\documentclass[dvips]{article}
\usepackage{supertabular,lscape,epsfig}
\usepackage{amssymb}
\usepackage{amsmath}
\usepackage{rotating}
\usepackage{booktabs}
\DeclareSymbolFont{ppa}{OT1}{ppl}{m}{it}
\DeclareMathSymbol{\vv}{\mathalpha}{ppa}{'166}

\begin{document}

\newcommand{\dd}{\,{\rm d}}
\newcommand{\ie}{{\it i.e.},\,}
\newcommand{\etal}{{\it et al.\ }}
\newcommand{\eg}{{\it e.g.},\,}
\newcommand{\cf}{{\it cf.\ }}
\newcommand{\vs}{{\it vs.\ }}
\newcommand{\zdot}{\makebox[0pt][l]{.}}
\newcommand{\up}[1]{\ifmmode^{\rm #1}\else$^{\rm #1}$\fi}
\newcommand{\dn}[1]{\ifmmode_{\rm #1}\else$_{\rm #1}$\fi}
\newcommand{\upd}{\up{d}}
\newcommand{\uph}{\up{h}}
\newcommand{\upm}{\up{m}}
\newcommand{\ups}{\up{s}}
\newcommand{\arcd}{\ifmmode^{\circ}\else$^{\circ}$\fi}
\newcommand{\arcm}{\ifmmode{'}\else$'$\fi}
\newcommand{\arcs}{\ifmmode{''}\else$''$\fi}
\newcommand{\MS}{{\rm M}\ifmmode_{\odot}\else$_{\odot}$\fi}
\newcommand{\RS}{{\rm R}\ifmmode_{\odot}\else$_{\odot}$\fi}
\newcommand{\LS}{{\rm L}\ifmmode_{\odot}\else$_{\odot}$\fi}
\newcommand{\feh}{\hbox{$ [{\rm Fe}/{\rm H}]$}}

\newcommand{\Abstract}[2]{{\footnotesize\begin{center}ABSTRACT\end{center}
\vspace{1mm}\par#1\par
\noindent
{~}{\it #2}}}

\newcommand{\TabCap}[2]{\begin{center}\parbox[t]{#1}{\begin{center}
  \small {\spaceskip 2pt plus 1pt minus 1pt T a b l e}
  \refstepcounter{table}\thetable \\[2mm]
  \footnotesize #2 \end{center}}\end{center}}

\newcommand{\TableSep}[2]{\begin{table}[p]\vspace{#1}
\TabCap{#2}\end{table}}

\newcommand{\FigCap}[1]{\footnotesize\par\noindent Fig.\  %
  \refstepcounter{figure}\thefigure. #1\par}

\newcommand{\TableFont}{\footnotesize}
\newcommand{\TableFontIt}{\ttit}
\newcommand{\SetTableFont}[1]{\renewcommand{\TableFont}{#1}}

\newcommand{\MakeTable}[4]{\begin{table}[htb]\TabCap{#2}{#3}
  \begin{center} \TableFont \begin{tabular}{#1} #4
  \end{tabular}\end{center}\end{table}}

\newcommand{\MakeTableSep}[4]{\begin{table}[p]\TabCap{#2}{#3}
  \begin{center} \TableFont \begin{tabular}{#1} #4
  \end{tabular}\end{center}\end{table}}

\newenvironment{references}%
{
\footnotesize \frenchspacing
\renewcommand{\thesection}{}
\renewcommand{\in}{{\rm in }}
\renewcommand{\AA}{Astron.\ Astrophys.}
\newcommand{\AAS}{Astron.~Astrophys.~Suppl.~Ser.}
\newcommand{\ApJ}{Astrophys.\ J.}
\newcommand{\ApJS}{Astrophys.\ J.~Suppl.~Ser.}
\newcommand{\ApJL}{Astrophys.\ J.~Letters}
\newcommand{\AJ}{Astron.\ J.}
\newcommand{\IBVS}{IBVS}
\newcommand{\PASJ}{PASJ}
\newcommand{\PASP}{P.A.S.P.}
\newcommand{\Acta}{Acta Astron.}
\newcommand{\MNRAS}{MNRAS}
\renewcommand{\and}{{\rm and }}
\section{{\rm REFERENCES}}
\sloppy \hyphenpenalty10000
\begin{list}{}{\leftmargin1cm\listparindent-1cm
\itemindent\listparindent\parsep0pt\itemsep0pt}}%
{\end{list}\vspace{2mm}}

\def\TYLDA{~}
\newlength{\DW}
\settowidth{\DW}{0}
\newcommand{\dw}{\hspace{\DW}}

\newcommand{\refitem}[5]{\item[]{#1} #2%
\def\REFARG{#3}\ifx\REFARG\TYLDA\else, {\it#3}\fi
\def\REFARG{#4}\ifx\REFARG\TYLDA\else, {\bf#4}\fi
\def\REFARG{#5}\ifx\REFARG\TYLDA\else, {#5}\fi.}

\newcommand{\Section}[1]{\section{#1}}
\newcommand{\Subsection}[1]{\subsection{#1}}
\newcommand{\Acknow}[1]{\par\vspace{5mm}{\bf Acknowledgments.} #1}
\pagestyle{myheadings}

\newfont{\bb}{ptmbi8t at 12pt}
\newcommand{\xrule}{\rule{0pt}{2.5ex}}
\newcommand{\xxrule}{\rule[-1.8ex]{0pt}{4.5ex}}
\def\thefootnote{\fnsymbol{footnote}}

\begin{center}
{\Large\bf Blue Large-Amplitude Pulsators and Other Short-Period Variable Stars in OGLE-IV Fields of the Outer Galactic Bulge}
\vskip1cm
{\bf
J.~~B~o~r~o~w~i~c~z$^1$,~~P.~~P~i~e~t~r~u~k~o~w~i~c~z$^1$,~~J.~~S~k~o~w~r~o~n$^1$,~~I.~~S~o~s~z~y~\'n~s~k~i$^1$,\\
A.~~U~d~a~l~s~k~i$^1$,~~M.~K.~~S~z~y~m~a~\'n~s~k~i$^1$,~~K.~~U~l~a~c~z~y~k$^2$,\\
R.~~P~o~l~e~s~k~i$^1$,~~S.~~K~o~z~{\l}~o~w~s~k~i$^1$,~~P.~~M~r~\'o~z$^1$,\\
D.~M.~~S~k~o~w~r~o~n$^1$,~~K.~~R~y~b~i~c~k~i$^{1,3}$,~~P.~~I~w~a~n~e~k$^1$,\\
M.~~W~r~o~n~a$^1$,~~M.~~G~r~o~m~a~d~z~k~i${^1}$,~~and~~M.~J.~~M~r~\'o~z$^1$\\}
\vskip3mm
{${}^1$ Astronomical Observatory, University of Warsaw, Al. Ujazdowskie 4, 00-478 Warszawa, Poland \\
${}^2$ Department of Physics, University of Warwick, Coventry CV4 7AL, UK \\
${}^3$ Department of Particle Physics and Astrophysics, Weizmann Institute of Science, Rehovot 76100, Israel \\}
\end{center}

\Abstract{In this work, we search the OGLE-IV outer Galactic bulge fields for short-period variable objects. The investigation focuses on unexplored timescales roughly below one hour in an area containing about 700 million stellar sources down to $I\approx20$~mag. We concentrate mainly on Blue Large-Amplitude Pulsators (BLAPs), which represent a recently discovered enigmatic class of short-period hot subluminous stars. We find 33 BLAPs in the period range from 7.5 to 66.5~min. Thirty-one of them are new discoveries, which increases the number of known stars of this class to over one hundred. Additional eighteen objects with pulsation-like light curve shapes and periods ranging from 17.3 to 53.7~min are presented. Very likely, these are $\delta$~Sct/SX~Phe-type stars, but some of them could be pulsating hot subdwarfs or BLAPs. We also report on the detection of five eclipsing binary systems with orbital periods between 61.2 and 121.9~min and three mysterious variable objects with {\it I}-band amplitudes larger than 0.9~mag. A spectroscopic follow-up would help in the final classification of the variables.}

{Stars: oscillations (including pulsations) --- Stars: variables: Blue Large-Amplitude Pulsators --- binaries: eclipsing}

\section{Introduction}

Short-period variable stars constitute a broad and diverse sort of celestial objects exhibiting brightness variations on timescales below about one hour. Among them could be single pulsating stars of early spectral types as well as close binary systems formed of low-mass red dwarfs or brown dwarfs. In recent years, the number of known short-period variables has significantly increased thanks to large-scale photometric surveys. One of the latest discovered class of pulsating stars, known as Blue Large-Amplitude Pulsators (BLAPs), was identified using data from the Optical Gravitational Lensing Experiment (OGLE; Udalski \etal 2015). The first BLAP was found in the Galactic disk at longitude $l=288\zdot\arcd06$, while other thirteen objects were detected toward the Galactic bulge (Pietrukowicz \etal 2017). The original 14 pulsators exhibit a uniform class of periodic variables with pulsation periods ranging from about 20 to 40~min and {\it I}-band amplitudes between about 0.2 and 0.4~mag. The stars were recognizable by their characteristic sawtooth-like light curves. Spectral analysis indicated that BLAPs belong to hot evolved objects with effective temperatures around 30~000~K. However, subsequent studies suggested that BLAPs represent a more diverse group. Four objects characterized by higher surface gravity and shorter periods (below 8~min) were discovered in data from the Zwicky Transient Facility (ZTF) survey (Kupfer \etal 2019). Further discoveries (McWhirter and Lam, 2022) were made using combined data from ZTF (Bellm \etal 2019) and Gaia space mission (Gaia Collaboration \etal 2021). Three new objects were detected in the OmegaWhite (OW) survey (Ramsay \etal 2022). The OGLE data for the Galactic disk revealed 20 new BLAPs with periods ranging from 8 to 62 min (Borowicz \etal 2023), while data for the inner Galactic bulge allowed the identification of additional 23 objects with periods ranging from 14 to 74 min, some of which showing an extra bump in the light curve (Pietrukowicz \etal 2024). Very recently, a BLAP located on the outskirts of the bulge was detected (Chang \etal 2024) using data from the SkyMapper Southern Sky Survey (Onken \etal 2019). So far, the number of known BLAPs stands at approximately 80, with over 60 of them identified by the OGLE survey.

Precise boundaries of parameters characterizing BLAPs remain uncertain. Recent studies suggest that the variables can exhibit a wider range of periods and amplitudes than previously thought, including multi-periodicity (Pietrukowicz \etal 2024, Koen \etal 2024). Until recently no BLAPs in binary systems were known. Using Transisting Exoplanetary Survey Satelite (TESS) photometry, Pigulski \etal (2022) detected a pulsator with a period of 32.37 min orbiting a hot main-sequence (MS) star with a 23.08 d period. Another object, TMTS-BLAP-1, discovered by the Tsinghua University -- Ma Huateng Telescopes for Survey, likely constitutes a wide binary system with an orbital period of 1576 d (Lin \etal 2023a).

BLAPs, as a class of variable stars, have sparked several hypotheses regarding their nature and evolutionary pathways. One of the proposed hypotheses suggests that BLAPs represent low-mass stars in the pre-white dwarf (pre-WD) phase, where they generate energy through residual hydrogen burning atop a degenerate helium core (Romero \etal 2018). Alternative models propose that BLAPs may be stars undergoing helium burning in their cores (Byrne and Jeffery, 2018; Wu and Li, 2018), or they could be shell helium-burning subdwarfs (Xiong \etal 2022). Stability of the BLAP pulsation periods, characterized by a change rate of $|\dot P/P|<10^{-6}$ yr$^{-1}$, indicates that these objects evolve on nuclear timescales. Proposed mechanisms explaining the evolutionary trajectory of BLAPs include origin from an evolved  binary system, following a common envelope phase with Roche lobe overflow (Byrne \etal 2021), or a result of a helium WD+MS merger (Zhang \etal 2023). Additionally, it has been proposed that BLAPs could emerge as surviving companions of single-degenerate Type Ia supernovae (Meng \etal 2020). The study of BLAPs can give us an insight into the structure and physical processes in hot pulsators---stars that oscillate due to the presence of a Z-bump in the oppacity at temperatures of about 200~000 K (Byrne and Jeffery, 2018). 

Among the short-period variables, binary systems form a separate class of objects. While the vast majority of binary stars have orbital periods longer than 0.2~d, there are also systems with shorter periods. In AM CVn systems, a type of cataclysmic variable stars, hydrogen-poor matter from a compact companion is accreted by a white dwarf. Binary systems with orbital periods below 30~min are usually formed of two white dwarfs. In recent years, the detection of multiple compact binary systems has become possible thanks to large-scale and all-sky variability surveys like ZTF, TESS, and Gaia (\eg Burdge \etal 2019, Burdge \etal 2020, Barlow \etal 2022, Ren \etal 2023).

In this work, we utilize time-series data collected by the OGLE project for fields located in the outer Galactic bulge. These fields have been surveyed for variable stars such as Cepheids (Udalski \etal 2018), RR Lyrae-type stars (Soszy{\'n}ski \etal 2019), $\delta$~Sct-type stars (Soszy{\'n}ski \etal 2021), and Mira stars (Iwanek \etal 2022). Here, we search for objects exhibiting periodic variations on timescales shorter than one hour. This paper is divided into several sections. In Section 2, we give a brief overview of the OGLE survey and the data we collected. Section 3 presents the data analysis methods. The detected short-period variables are described in Section 4. We wrap up with a summary of our results in Section 5.

\section{Observations}

Our variability search utilizes the photometric data from the fourth phase of the OGLE project (OGLE-IV), which commenced in March 2010. OGLE has been a leading large-scale survey for over three decades. The project observes dense stellar regions of the southern sky with the 1.3-meter Warsaw telescope situated at the Las Campanas Observatory, Chile. The observatory is operated by the Carnegie Institution for Science. The primary regions monitored by OGLE are the inner Galactic bulge and the Magellanic Clouds. Additional shallow observations of the Galactic disk and outer Galactic bulge are conducted since 2013 as the \textit{Galaxy Variability Survey} (GVS), with a cadence of a few days over the first 100 epochs. Currently, the project covers an area of nearly 3600 square degrees. OGLE-IV encompasses 622 Galactic bulge (BLG) fields, 444 of which belong to the outer bulge and are distributed approximately evenly around the Galactic center. The OGLE-IV BLG fields cover an area spanning roughly from $-15^\circ$ to $+20^\circ$ in Galactic longitude and from $-15^\circ$ to $+15^\circ$ in Galactic latitude. Coordinates of the OGLE-IV BLG fields can be found at
\begin{center}
\textit{https://ogle.astrouw.edu.pl/sky/ogle4-BLG/}
\end{center}

The OGLE-IV project employs a 268.8-megapixel mosaic camera with a pixel size of $0\zdot\arcs26$ and a 1.4-square-degree field of view, consisting of 32 single 2K$\times$4K CCD detectors. Observations are conducted using {\it I} and {\it V} filters, which closely resemble those of the standard Johnson-Cousins system. Observations in the {\it I}-band are much more frequent than in the {\it V}-band. Exposure times in the {\it I}-band are of 25~s for the GVS fields and of 100~s for the inner bulge fields. Unfortunately, in the case of the outer bulge fields, regular {\it I}-band measurements (ranging from 90 to 250 epochs) are not supplemented with {\it V}-band measurements. The {\it I}-band brightness range in GVS fields spans from about 10.5 to 21~mag. An exception are seven fields, from BLG705 to BLG711, covering the central part of the Sagittarius dwarf spheroidal galaxy (SgrDG). The data for this small region were taken in both {\it V} and {\it I} bands with 150-s exposure times in years 2011--2014 and contain around 150 epochs in the {\it I}-band. The OGLE-IV observations were stopped on March 18, 2020, as a result of the COVID-19 pandemic, but regular monitoring resumed on August 12, 2022. Technical details on the survey can be found in Udalski \etal (2015).

\section{Data Analysis}

The search for short-period variable stars was conducted using time-series data in the {\it I}-band. We analyzed 444 OGLE-IV fields for the outer Galactic bulge area from the shallow survey and seven fields from the deep survey of the SgrDG, totalling about 700 million stellar sources. We focused on sources with more than 50 measurements and brighter than $I=20$ mag. Our initial step involved an identification of periodic signals for all the sources using the \textit{FNPEAKS}\footnote[1]{http://helas.astro.uni.wroc.pl/deliverables.php?active=fnpeaks\&lang=en} code, which employs a discrete Fourier transform on unevenly spaced data. We explored a frequency range of $0~<~f~<~500$ cycles per day ($P~>~2.88$~min). Subsequently, only objects exhibiting a significant signal-to-noise ratio were retained. The variability signal cutoff varied across different fields, mainly depending on the number of observations ($S/N \sim \sqrt{N_{\rm obs}^{}}$).

In the following stage, all variable objects with periods shorter than 0.05~d underwent a visual inspection. We refined our candidate variable lists by eliminating artifacts (objects displaying spurious variability). Any noticeable outliers in the light curves of genuine variables were removed for further analysis. About three hundred variable stars with periods below 60 min were found. Stars exhibiting amplitudes exceeding 0.05 mag underwent further investigation. BLAP candidates were identified based on the characteristic light curve morphology. Variables with low amplitudes and sinusoidal-like light curves could not be classified solely based on the photometric data. Most of these objects are likely low-amplitude $\delta$~Sct stars, similar to those found in several recent studies (\eg Pietrukowicz \etal 2022, Lin \etal 2023b), although among them there can be other short-period variables such as pulsating subdwarfs, SX~Phe-type stars, or low-inclination eclipsing binaries. Light curves exhibiting uneven minima upon doubling of the period were also selected for further inspection. Instrumental magnitudes were then converted to the standard photometric system. The accuracy of the calibration, which follows the method outlined in Udalski \etal (2015), achieves a precision of 0.02 mag within the Johnson-Cousins system.

As there were no available {\it V}-band measurements in the examined outer bulge fields, we constructed color-magnitude diagrams (CMDs) using data from Gaia Data Release 3 (Gaia Collaboration \etal 2021). The diagrams were prepared based on the mean {\it G}-band magnitudes and $(BP-RP)$ colors for stars located within a $0.1\times0.1$ square degrees region. For two objects without available $(BP-RP)$ measurements, data from the DECaPS DR2 (Saydjari \etal 2023) were used. To analyze the short-period variables, we transformed the brightness measurement timestamps from Heliocentric Julian Date (HJD) to the more precise Barycentric Julian Date (BJD$_{\rm TDB}$). The previously found periods were then corrected using the \textit{TATRY} code (Schwarzenberg-Czerny 1996) which utilizes multi-harmonic functions. The final list of BLAPs was compiled based on the morphology of the light curve, period, amplitude, and position on the CMD. All objects of interest were then examined for the presence of any additional periodicities. 

\section{Results}

The primary result of this work are the compilations of BLAPs, eclipsing binary systems, and other (unclassified) short-period variable objects detected within the OGLE-IV outer Galactic bulge area. Phase-folded light curves and CMDs for the discovered variables are presented in this paper, while time-series data and finding charts for BLAPs can be downloaded from the on-line version of the OGLE Collection of Variable Stars (OCVS) at
\begin{center}
\textit{https://www.astrouw.edu.pl/ogle/ogle4/OCVS/} 
\end{center}
For BLAPs, we adopt the same naming convention as for other variable stars in the OGLE collections. We start numbering the new objects as OGLE-BLAP-NNN from NNN=062 and continue with the increasing right ascension.

\subsection{BLAPs}

Our search led to the detection of 33 genuine BLAPs. Thirty-one of these objects are new discoveries. Observational parameters including coordinates, pulsation periods, magnitudes, and {\it I}-band amplitudes of the identified BLAPs are provided in Table 1. Parallaxes ($\varpi$), {\it G}-band magnitudes, and $(BP-RP)$ colors are sourced from the Gaia EDR3 catalog (Gaia Collaboration \etal 2021). Two of the detected stars were previously identified as BLAPs, namely OGLE-BLAP-086 as object OW J1826-2812 by Ramsay \etal (2022) and OGLE-BLAP-093 as SMSS J184506.82-300804.7 by Chang \etal (2024). Positions on the sky of BLAPs detected in our search and reported earlier (Pietrukowicz \etal 2017, 2024; Borowicz \etal 2023) are shown in Fig.~1. In the bulge area, all pulsators but object OGLE-BLAP-093 lie within latitudes $2^{\circ}<|b|<8^{\circ}$. The lack of stars close to the Galactic plane is very likely caused by high extinction.

\begin{figure}
\includegraphics[width=12.75cm]{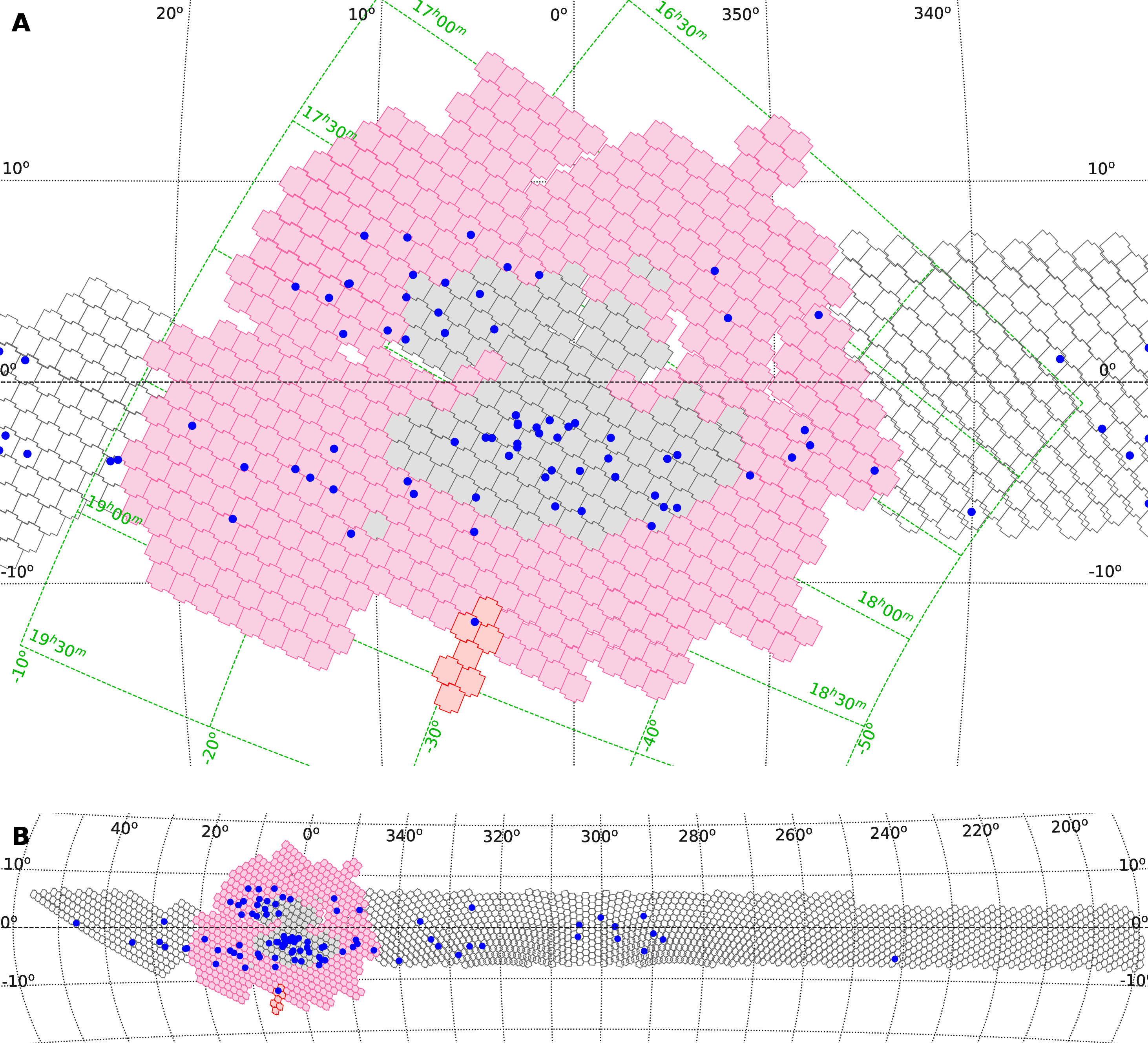}
\FigCap{Positions of BLAPs detected by the OGLE-IV survey in the Galactic bulge area (panel~A) and the whole Milky Way stripe monitored from Las Campanas Observatory (panel~B). The inner (deep) bulge fields are shown in gray, whereas the outer (shallow) bulge fields are in pink. Special (deep) fields covering the central part of the Sagittarius dwarf spheroidal galaxy are marked in red. The OGLE-IV Galactic disk fields are depicted in white. All BLAPs but one object lie within the galactic latitudes $|b|<8^{\circ}$. The lack of BLAPs at latitudes $|b|<2^{\circ}$ in the bulge area is likely due to high extinction.}
\end{figure}

In Fig.~2, we present the period vs. {\it I}-band amplitude diagram, depicting the positions of all BLAPs identified by the OGLE survey, including stars detected in this study. Additionally, the figure includes single-mode $\delta$~Sct-type stars from the OCVS (Soszy{\'n}ski \etal 2021). Notably, there are no obvious differences in the location on the diagram between objects observed in the disk and bulge fields. All BLAPs with periods shorter than 70~min have the amplitudes larger than 0.1~mag. The only exception is object OGLE-BLAP-022 with a period of 74.05~min (the longest known so far) and an amplitude of 0.087 mag in the {\it I}-band (Pietrukowicz \etal 2024). The highest measured amplitude in {\it I} is 0.428 mag in the case of OGLE-BLAP-078. It can be noted that the upper bound on the amplitude decreases with the increasing pulsation period.

Four out of the 33 identified BLAPs have been classified as variable objects in the Gaia DR3, namely OGLE-BLAP-069, 071, 091, and 094. These objects have been assigned to the DSCT$|$GDOR$|$SXPHE class. Phase-folded {\it I}-band light curves and {\it I} \vs $V-I$ CMDs for 32 BLAPs detected in the OGLE-IV outer Galactic bulge fields are shown in Figs.~3--9. In the case of object OGLE-BLAP-082, a CMD is not provided as it does not have the $(BP-RP)$ color in the Gaia data, and DECaPS DR2 does not cover the vicinity of this star. Instead, in Fig.~10, we present a color image of a nearby region from Pan-STARRS DR1 (Chambers \etal 2016). 
\begin{centering}
\begin{figure}
\centering
\includegraphics[width=10.5cm]{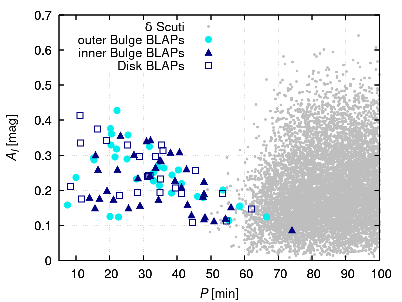}
\FigCap{Period \vs {\it I}-band amplitude diagram for all 94 BLAPs detected in the OGLE-IV fields. Positions of the 33 BLAPs from the outer Galactic bulge presented in this work are marked with filled cyan circles. Twenty-five stars from the Galactic disk fields (including the prototype object OGLE-BLAP-001) are shown with blue empty squares. Filled blue triangles mark the positions of 36 BLAPs found in the inner Galactic bulge fields. $\delta$~Sct-type stars from the OGLE fields for the Milky Way (bulge and disk) are represented with gray points.}
\end{figure}
\end{centering}
Some of the detected objects exhibit blur in their light curves, which is likely caused by period changes. This can be seen in the case of OGLE-BLAP-085 and OGLE-BLAP-091. We note that the majority of observations for these stars comes from seasons 2013, 2014, and 2023. The discovered BLAPs exhibit characteristic light curve shapes, blue colors, and large amplitudes. The periods of the identified objects range from 7.5 to 66.5~min and the {\it I}-band amplitudes range from 0.11 to 0.43~mag. Object OGLE-BLAP-089, with a period of 7.51~min, supersedes OGLE-BLAP-044 ($P$~=~8.42~min) as the shortest BLAP discovered in the OGLE data. Among the presented objects, OGLE-BLAP-069, has the longest period of 66.49~min. Several distinct features can be recognized in the phased light curves of the identified objects, with the majority of objects showing classical sawtooth-like shape. The next group of light curves is characterized by a sinusoidal-like bump after the maxiumum lasting for about half of the cycle. A similar effect is observed in extreme helium stars BX Cir ($P$~=~153.6~min) and V652 Her ($P$~=~155.4~min) (Kilkenny \etal 1999), although these stars have significantly smaller amplitudes. Examples of objects exhibiting this effect are OGLE-BLAP-071 and OGLE-BLAP-083. Similar stars were also found in the Galactic disk (\eg OGLE-BLAP-043; Borowicz \etal 2023) and in the inner bulge fields (\eg OGLE-BLAP-019; Pietrukowicz \etal 2024). Light curves of some BLAPs show an inverted shape, where the rising branch lasts longer than the fading one (OGLE-BLAP-069 and OGLE-BLAP-085 in our sample), similar to the effect observed in OGLE-BLAP-022 from the inner bulge area. Another noticeable feature in some light curves is a characteristic dip near the maximum brightness. This effect is visible in object OGLE-BLAP-009 in an inner Galactic bulge field (Pietrukowicz \etal 2017) and OGLE-BLAP-040 in a Galactic disk field (Borowicz \etal 2023). The additional features are more common for longer period objects, particularly for periods longer than 50 min.

Based on the Gaia EDR3 parallaxes (Gaia Collaboration \etal 2021), we can conclude that all the BLAPs reported here are located at distances of several kpc from the Sun, similarly to previously found objects. We note that there are no additional periodic signals down to amplitudes of 0.02 mag in the detected stars, although this can result from a relatively low number of brightness measurements.

\subsection{Other pulsation-like variables}

During the search, we found eighteen short-period variables with light curve shapes resembling pulsating stars, although the asymmetry is less pronounced than in the case of BLAPs. Figs.~11--14 shows phase-folded light curves and CMDs with positions of the variables. Since there are no available $(BP-RP)$ measurements in the Gaia database, we used {\it g} and {\it i}-band data obtained by the DECam Plane Survey to construct the CMDs for BLG882.04.34201 and BLG918.29.51389. We note that we did not find any other periodicities in the variables. For object BLG688.16.858, a distinct change in the period over the course of the observations can be noted, as its light curve is blurred. All the reported variables are located blueward of the main sequence. In Table~2, we list the observational properties of the stars ordered with the increasing right ascension. The parallax values around and below 1~mas and {\it I}-band brightnesses in range 15.5-19~mag indicate that the objects are not pulsating white dwarfs. The variables are very likely of $\delta$~Sct/SX~Phe type. However, some of them can be pulsating subdwarfs of spectral types OB (sdOB). We cannot rule out a possibility that there are BLAPs in this sample. Only object BLG481.23.763 was assigned a variability class in the Gaia DR3 (as DSCT$|$GDOR$|$SXPHE).

Interestingly, variable BLG954.25.16264 is located at an angular distance of only 4\zdot\arcm69 from the center of globular cluster NGC 6541. This is about 4.42 times its half-light radius ($r_{\rm h}=1$\zdot\arcm06, according to the 2010 edition of a catalog of parameters for Milky Way's globular clusters prepared by Harris 1996). On the CMD, the star falls in the area where the cluster blue stragglers reside. Very likely, object BLG954.25.16264 is a SX~Phe-type variable that belongs to NGC 6541.

\subsection{Eclipsing Binary Systems}

Our analysis of the OGLE-IV outer Galactic bulge dataset reveals five short-period eclipsing binary systems. These findings are illustrated in Fig.~15, where we present a CMD alongside the folded light curve for each system according to its orbital period. Furthermore, Table~3 provides detailed photometric properties for the objects. Only one of them, BLG764.18.16589, was classified as an eclipsing variable (ECL type) in the Gaia DR3, but with an incorrect period. Systems BLG723.31.34651 and BLG942.13.1292 may belong to the class of recently discovered Roche lobe-filling hot subdwarf binaries (Kupfer \etal 2020b). Their properties resemble those of objects ZTF J2130+4420 (Kupfer \etal 2020a) and ZTF J2055+4651 (Kupfer \etal 2020b), which are confirmed sdOB+WD binaries. Object BLG764.18.16598 with the period of 82.43~min is likely a HW Vir-type system with a strong reflection effect. This star was previously found as ZTF J1810-2138 and classified as a detached binary (Ren \etal 2023). System BLG608.02.2325 with the orbital period of 61.23~min could belong to one of the aforementioned types.

An intriguing object is BLG832.03.1534 with the orbital period of 121.86 min. The shape of the phased light curve with the minima of similar depth points to a contact binary system. Although currently, shorter-period hot contact binaries are known, the location of this system on the CMD suggests that it does not contain a hot subdwarf. Contact binaries with orbital periods below 0.2~d and formed of red dwarfs are very rare. Example systems at the short-period end are WFCAM 19b-3-06008 with $P_{\rm orb}\approx161$~min (Nefs \etal 2012) and OGLE-BLG-ECL-000066 with $P_{\rm orb}\approx141$~min (Soszy{\'n}ski \etal 2015), making BLG832.03.1534 a good candidate for the shortest known cool contact binary. With a parallax of 5.6~mas, the object is located relatively close to the Sun compared to the remaining binary systems reported here, thus it is likely composed of M dwarfs.

\subsection{Mysterious short-period variables}

Our search led to the detection of the following three unusual variables shown in Fig.~16. Object BLG696.13.32988 has a huge amplitude of about 1.6 mag at the extremely short period of 11.20~min. Its light curve seems to be symmetric. We inform that this star is a faint but isolated object in the OGLE images. The other two variables, BLG704.31.1645 and BLG391.22.1563, have very bizarre light curve shapes at a period of about 90~min. Their properties are very similar to one of the previously discovered objects in the OGLE-IV Galactic disk fields, GD2117.15.7515 with the period of $P=90.74$~min (Borowicz \etal 2023). Object BLG391.22.1563 with the huge {\it I}-band amplitude of about 2.5 mag was classified as a cataclysmic variable (CV) in Gaia DR3, which is a possible explanation for its nature. However, none of the objects reported in this work has an X-ray counterpart.

\section{Summary}

The presented study focused on the search for short-period ($<$~1~h) variable stars in the OGLE-IV outer Galactic bulge fields. We put a specific emphasis on the detection of BLAPs and compact binary systems. The monitoring of about 700~million stellar sources allowed us to detect around three hundred periodic variable sources, among which 33 are genuine BLAPs and five are eclipsing binary systems. Eighteen other of the detected large-amplitude ($>$~0.1~mag) variables are likely pulsating stars. We also found three variables of an uncertain type that have a huge amplitude ($>$~0.9~mag). The remaining detected objects are low-amplitude variables of unassigned type. 

Our investigation have elevated the number of BLAPs from OGLE to 94, bringing the total number of these stars to over one hundred. OGLE-BLAP-092 with a pulsation period of 7.51~min has replaced OGLE-BLAP-044 with $P$~=~8.42~min as the shortest-period BLAP discovered by OGLE. Some of the newly identified objects exhibit unusual features such as bumps or dips in their light curves. The light curves of objects OGLE-BLAP-085, OGLE-BLAP-091, and OGLE-BLAP-093 show noticeable alterations in their shape, which could be caused by a period changes. The substantial increase in the population of BLAPs allows us to put a better constraint on their parameters and provides targets for further investigations.

In addition to the BLAPs, we identified five short-period eclipsing binary systems with orbital periods in the range 61--121~min. These objects likely contain typical blue reflection-effect binaries of HW Vir type and compact sdB systems with ellipsoidal modulation, whereas objects BLG832.03.1534 is unusually red and has a very short period for a possible EW-type binary system. Two of the objects in our sample could be cataclysmic variables. Eighteen other variables with pulsation-like light curves could be BLAPs, short-period $\delta$~Sct stars or oscillating subdwarfs. Photometry alone is insufficient to determine the class of these variables. Follow-up spectroscopic observations would allow their final classification.

\begin{figure}
\includegraphics[width=12.71cm]{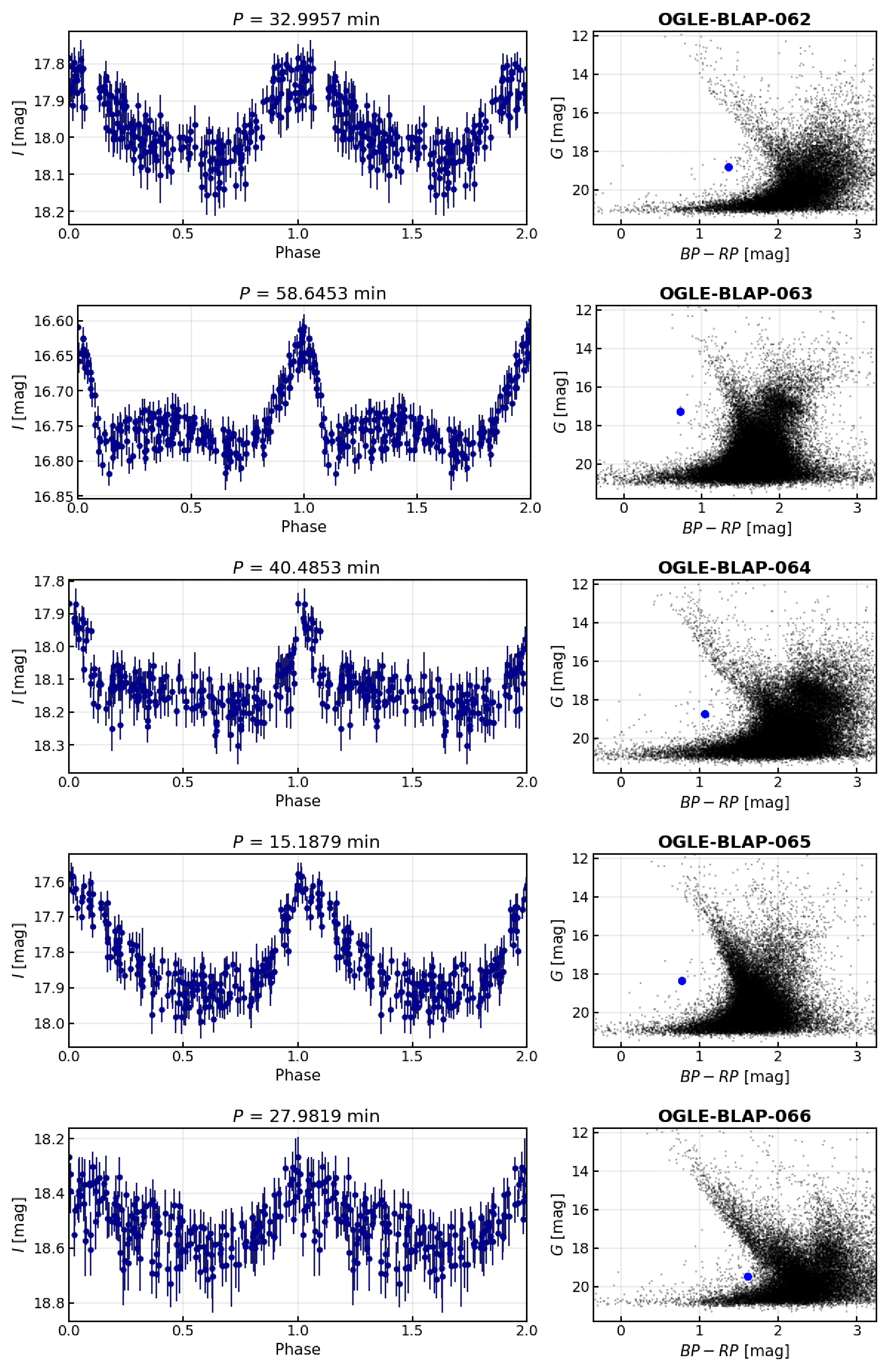}
\FigCap{Phase-folded {\it I}-band light curves (left panels) and CMDs (right panels) for BLAPs detected in the OGLE-IV fields for the outer Galactic bulge. The diagrams were constructed using data from Gaia DR3 within $0.1 \times 0.1$ square-degree regions centered on the variable objects. Positions of the BLAPs in the CMDs are marked with blue dots. Note the presence of an additional bump in the light curve of object OGLE-BLAP-063.}
\end{figure}

\begin{figure}
\includegraphics[width=12.71cm]{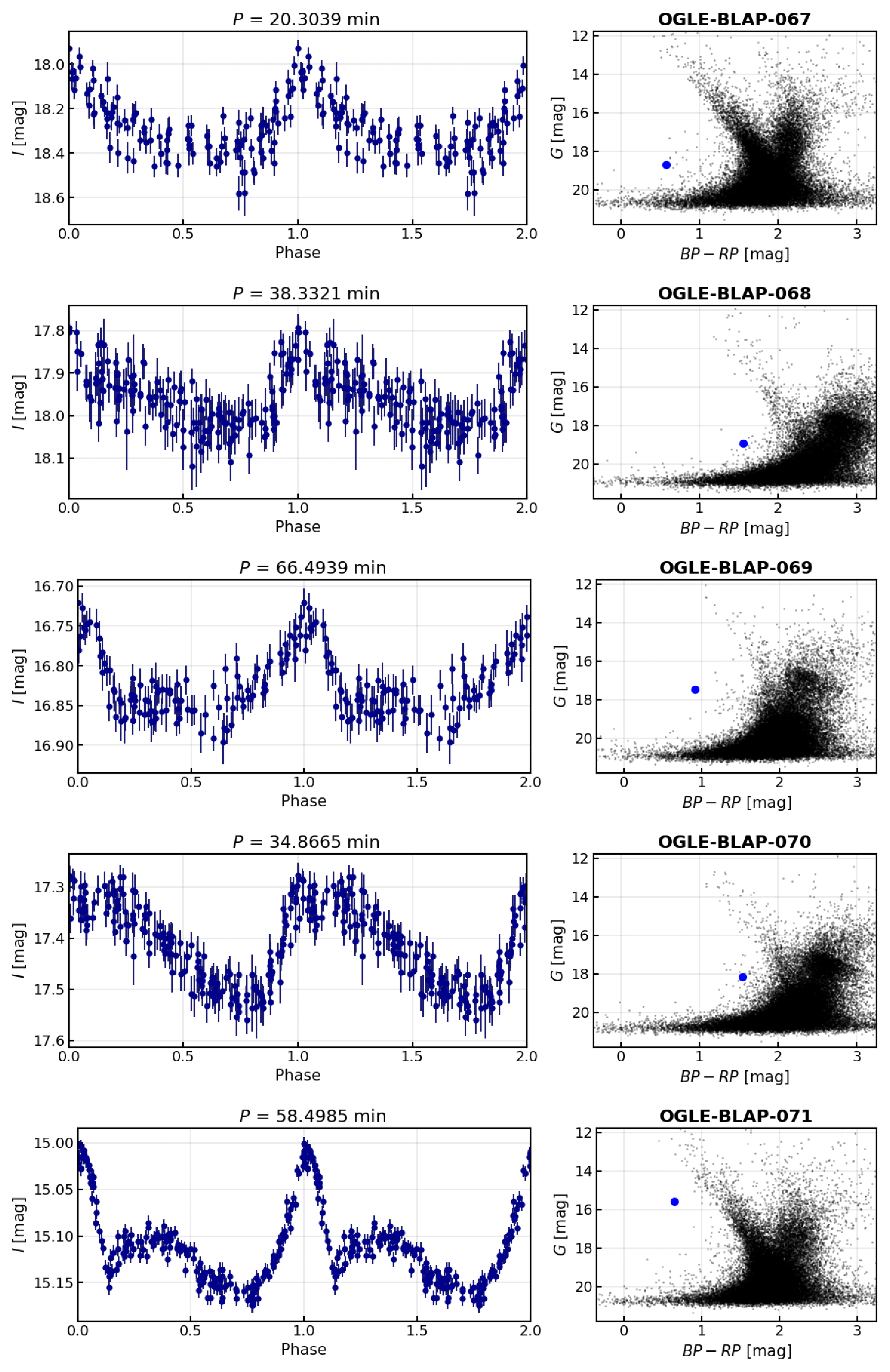}
\FigCap{Same as Fig.~3 for another five BLAPs. Object OGLE-BLAP-070 exhibits a characteristic dip around the maximum light, which is present in several other BLAPs including the best-studied case so far, OGLE-BLAP-009 (Bradshaw \etal 2024).}
\end{figure}

\begin{figure}
\includegraphics[width=12.71cm]{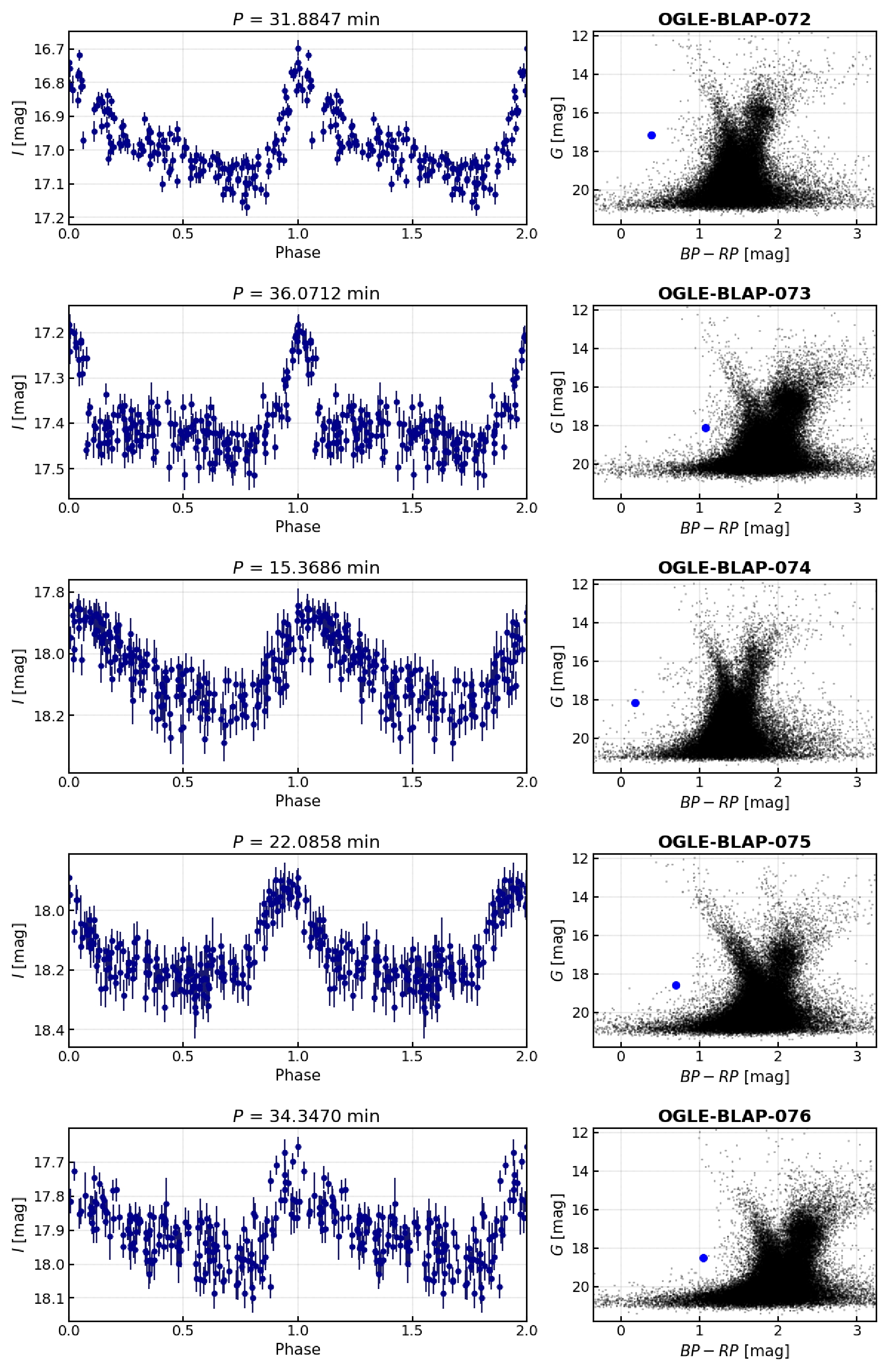}
\FigCap{Same as Fig.~3 for another five BLAPs.}
\end{figure}

\begin{figure}
\includegraphics[width=12.71cm]{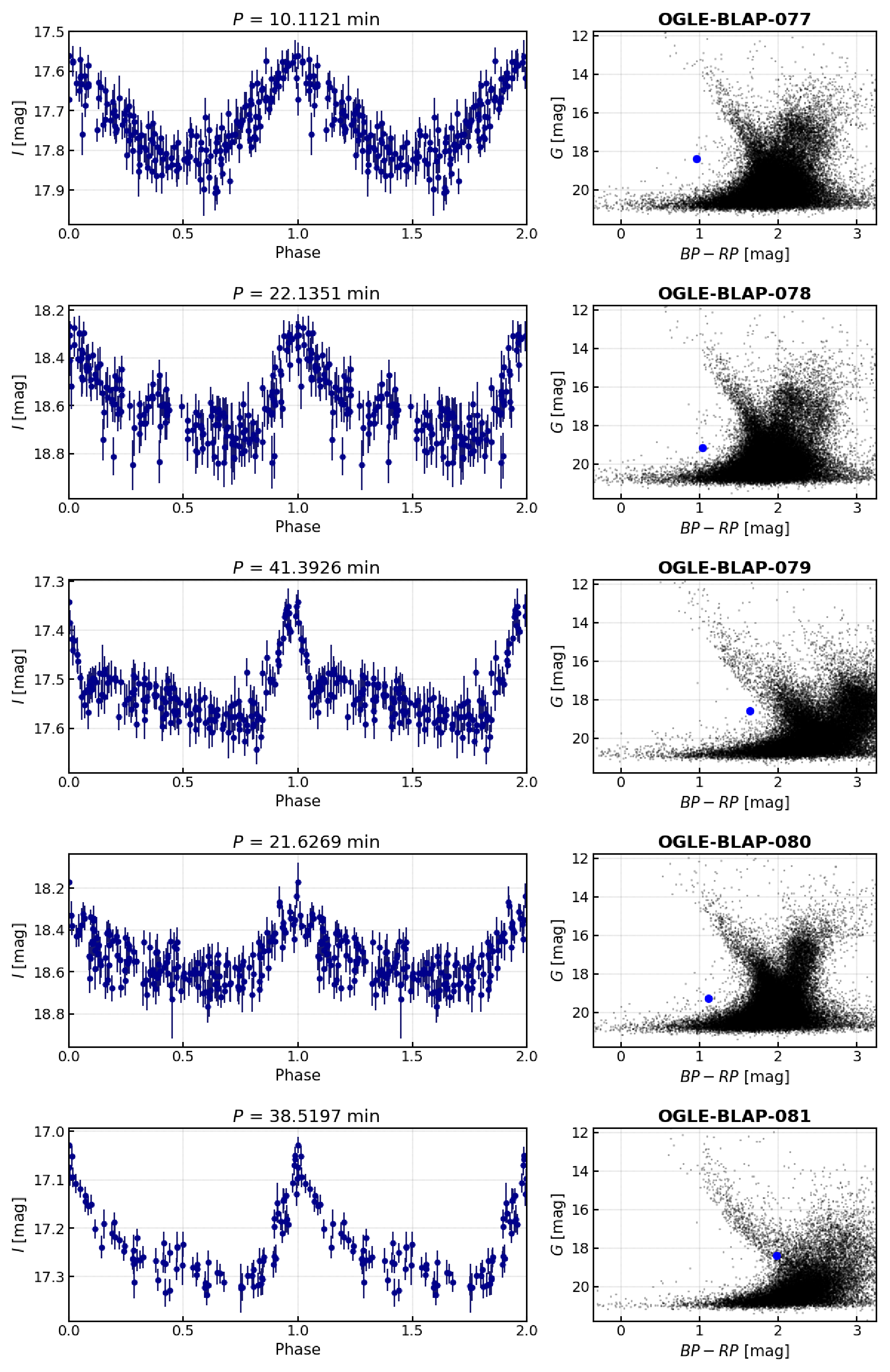}
\FigCap{Same as Fig.~3 for another five BLAPs.}
\end{figure}

\begin{figure}
\includegraphics[width=12.71cm]{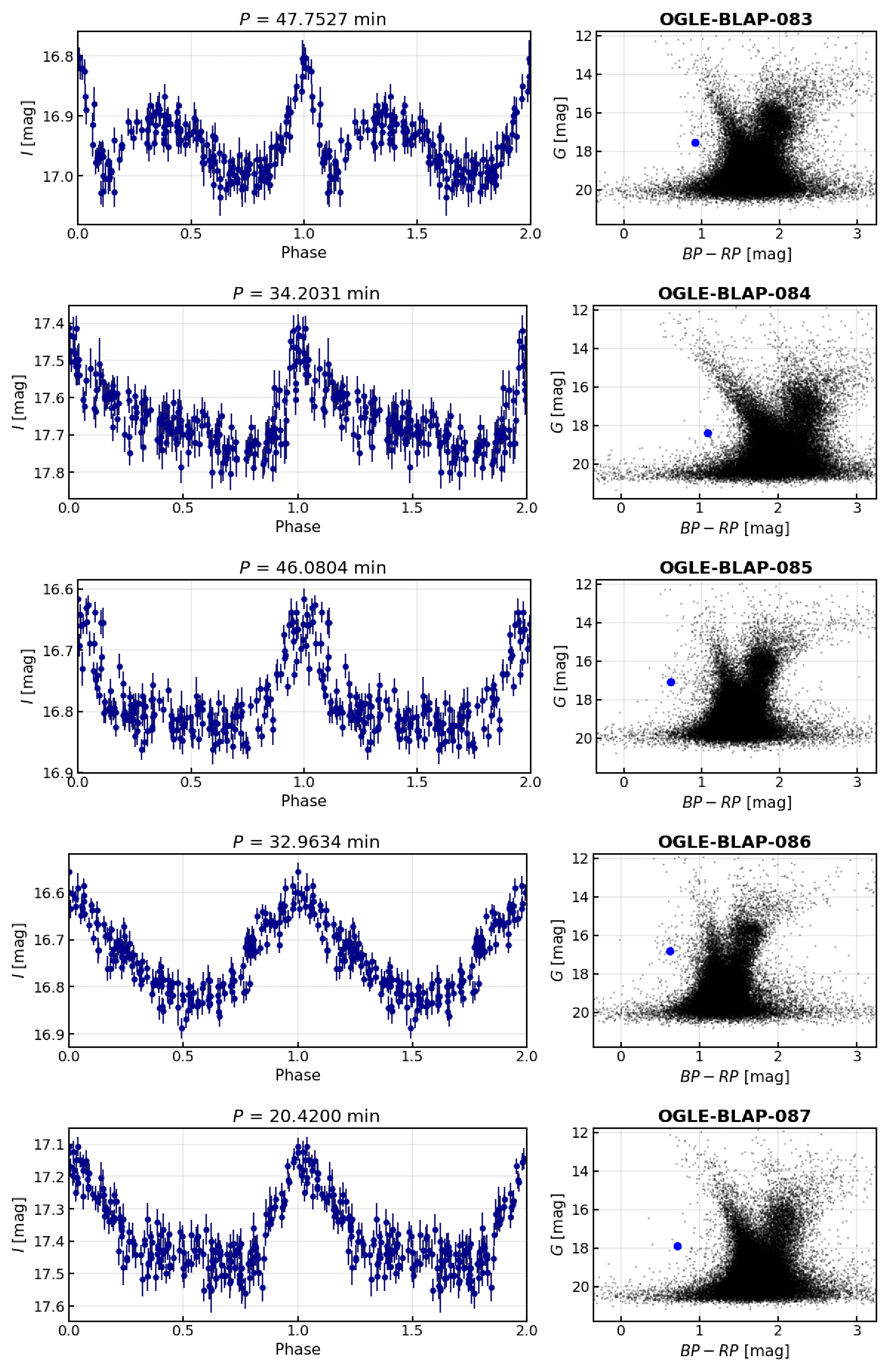}
\FigCap{Same as Fig.~3 for another five BLAPs. A prominent additional bump is present in the light curve of OGLE-BLAP-083.}
\end{figure}

\begin{figure}
\includegraphics[width=12.71cm]{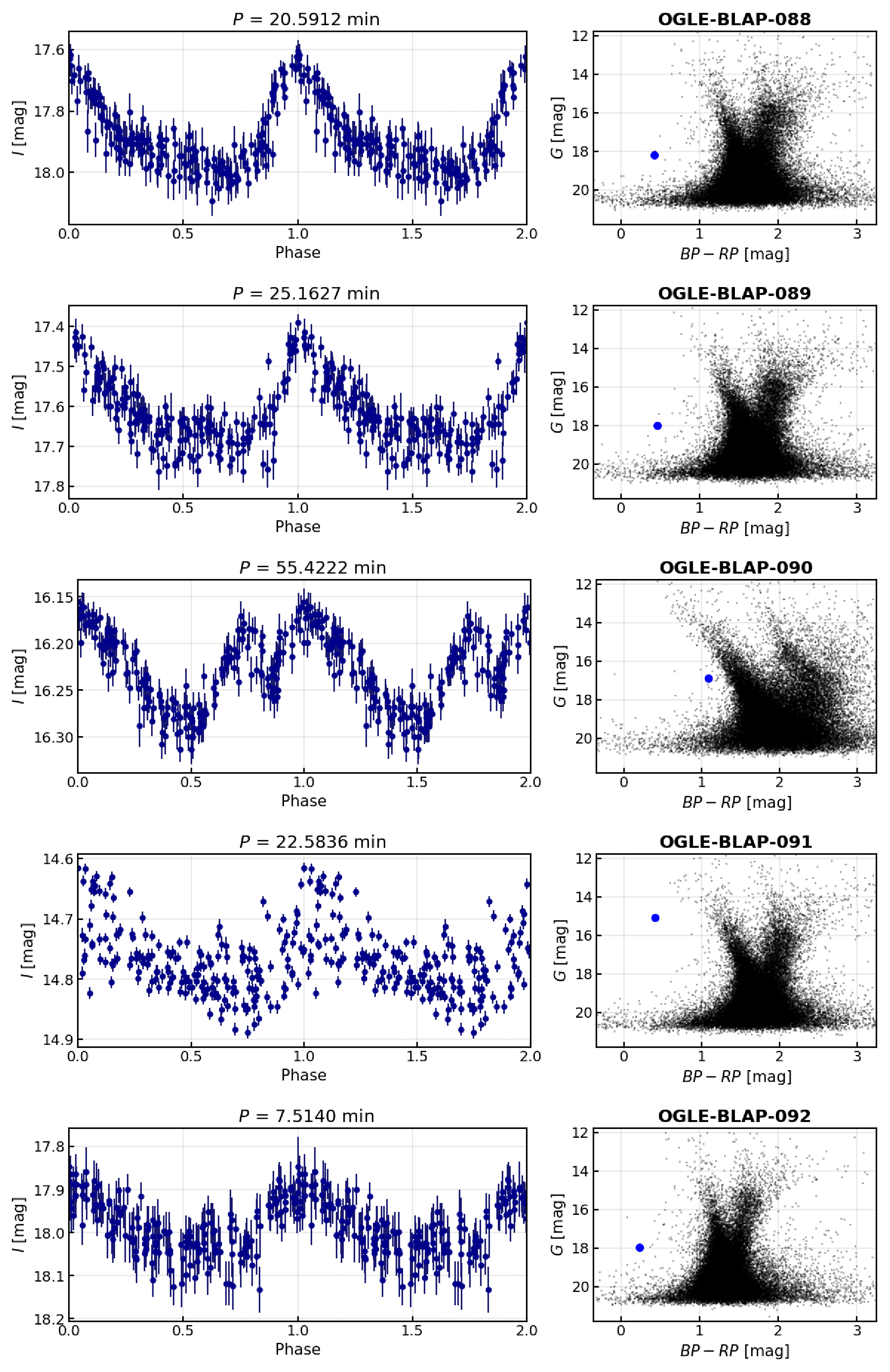}
\FigCap{Same as Fig.~3 for another five BLAPs. In the case of OGLE-BLAP-091, the blurred light curve is a result of period variations over a decade-long observations. Object OGLE-BLAP-092 has the shortest pulsation period in the whole BLAP collection from OGLE.}
\end{figure}

\begin{figure}
\includegraphics[width=12.71cm]{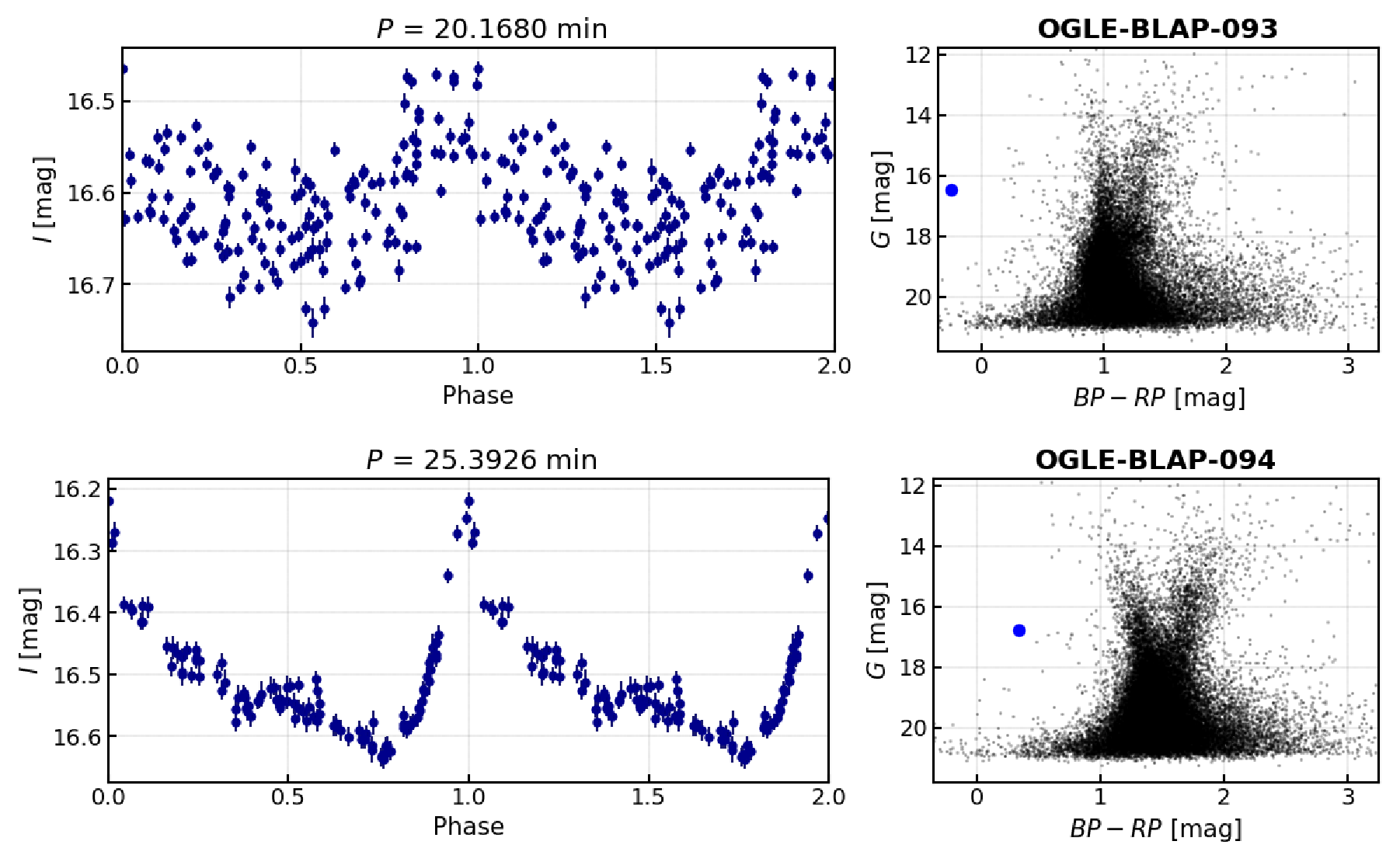}
\FigCap{Phase-folded {\it I}-band light curves along with Gaia DR3 CMDs for two remaining BLAPs. An unusual scatter can be observed in the light curve of OGLE-BLAP-093.}
\end{figure}

\begin{figure}
\includegraphics[width=12.7cm]{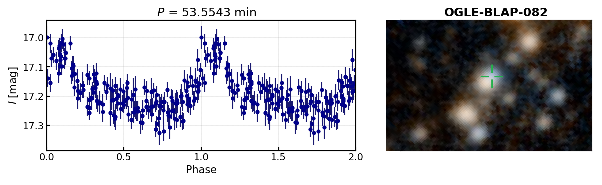}
\FigCap{Phase-folded {\it I}-band light curve of OGLE-BLAP-082 (left panel) and a 22"$\times$14" color chart cropped from a Pan-STARRS image with this variable star (right panel). Position of the star is marked with a green cross.}
\end{figure}

\begin{figure}
\includegraphics[width=12.71cm]{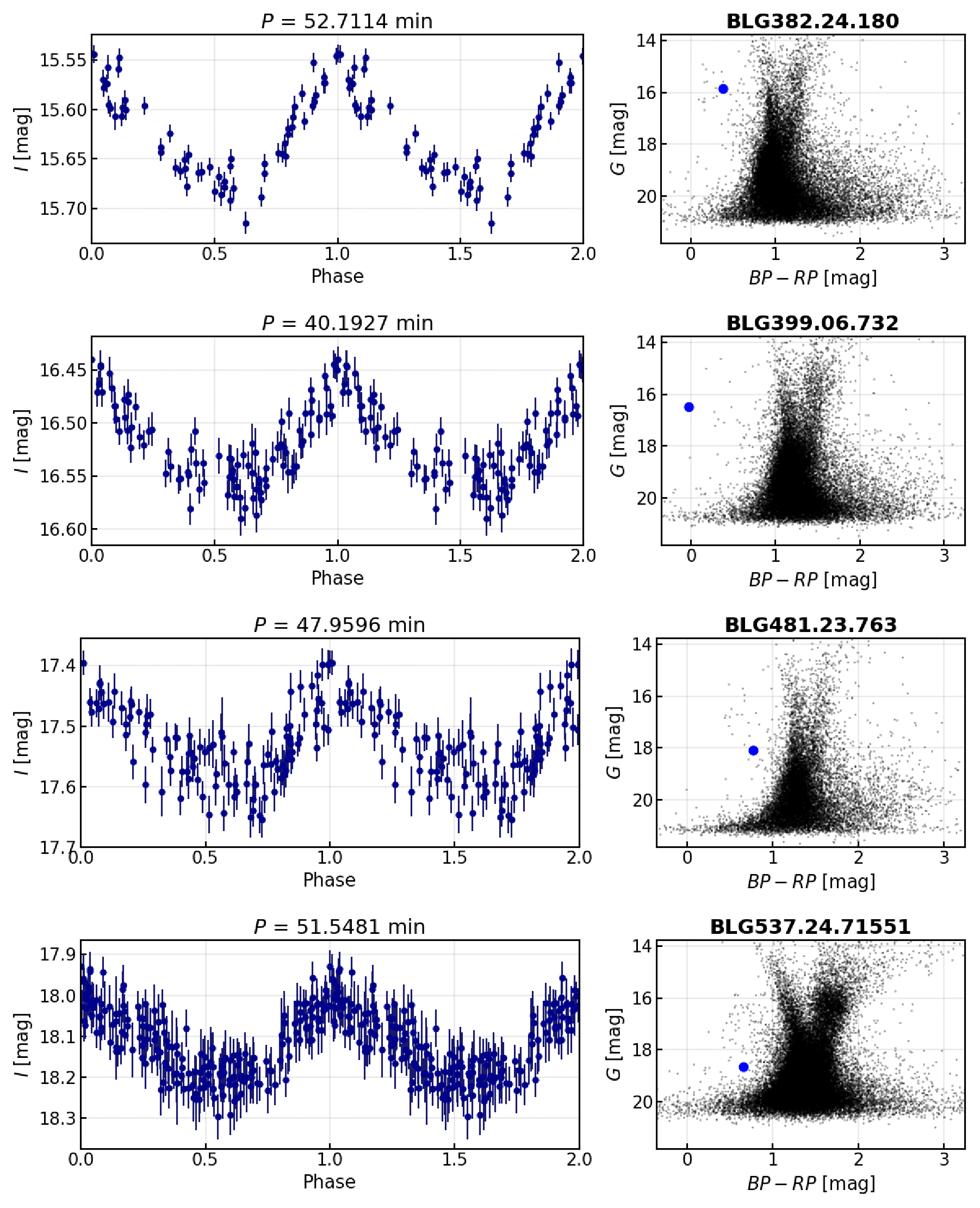}
\FigCap{Phase-folded {\it I}-band light curves (first column) and CMDs (second column) for four pulsation-like variable stars detected in the OGLE-IV fields for the outer Galactic bulge.}
\end{figure}
\begin{figure}
\includegraphics[width=12.71cm]{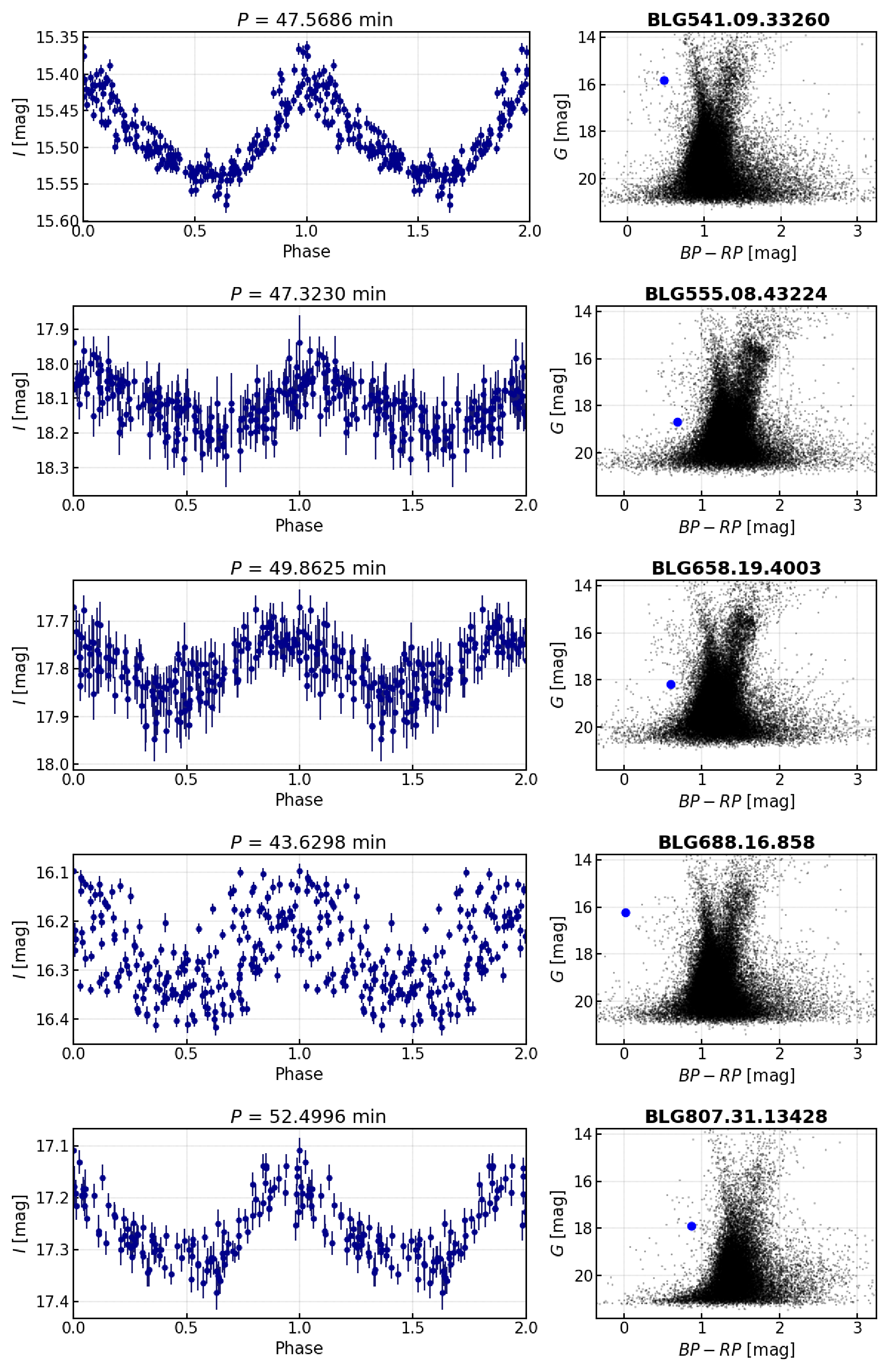}
\FigCap{Same as Fig.~11 for another five objects. In the case of object BLG688.16.858, a period change results in the blurred light curve.}
\end{figure}
\begin{figure}
\includegraphics[width=12.71cm]{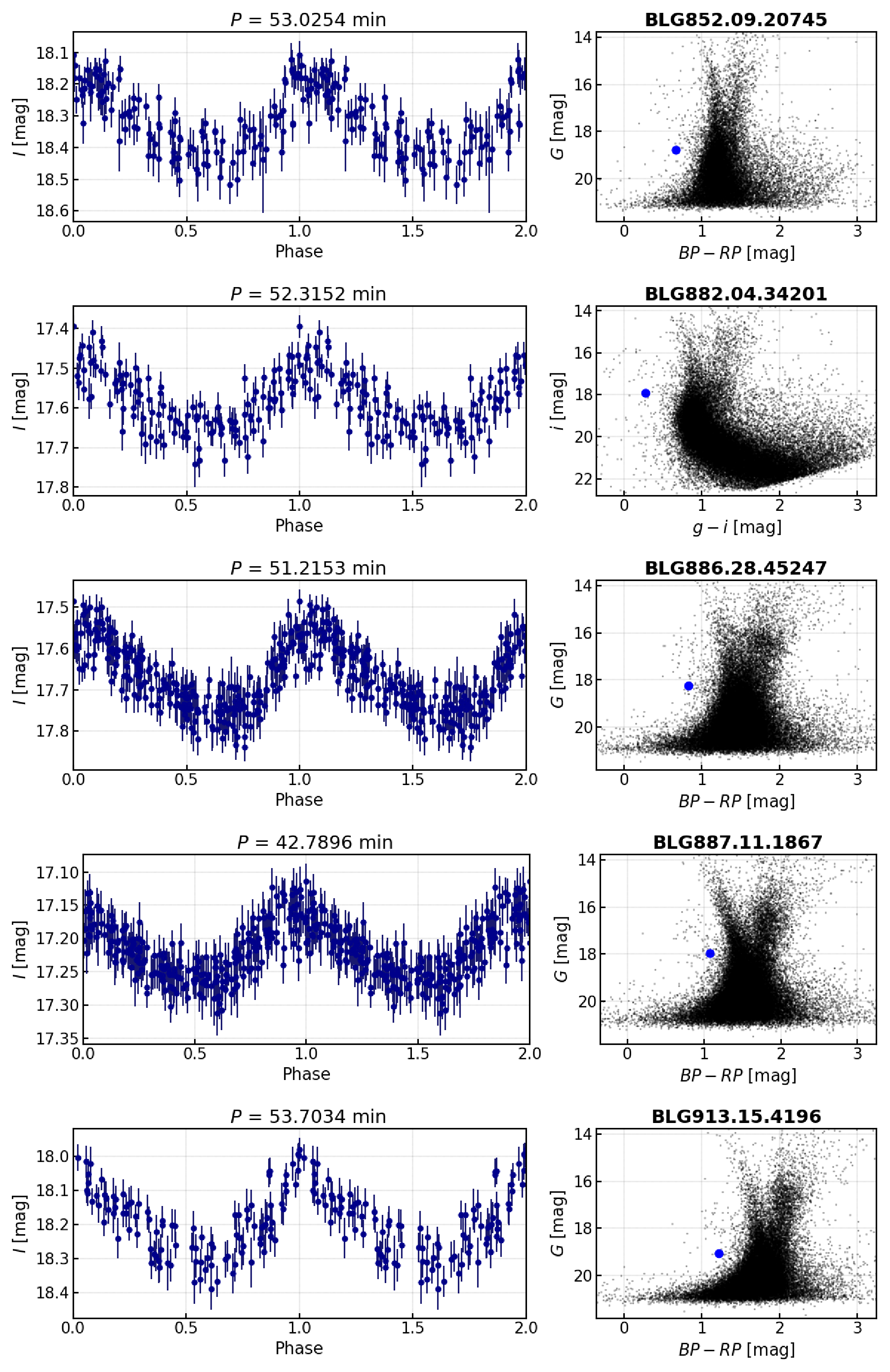}
\FigCap{Same as Fig.~11 for another five objects. The CMDs for BLG882.04.34201 is based on {\it g} and {\it i}-band brightness measurements obtained from the DECam Plane Survey.}
\end{figure}
\begin{figure}
\includegraphics[width=12.71cm]{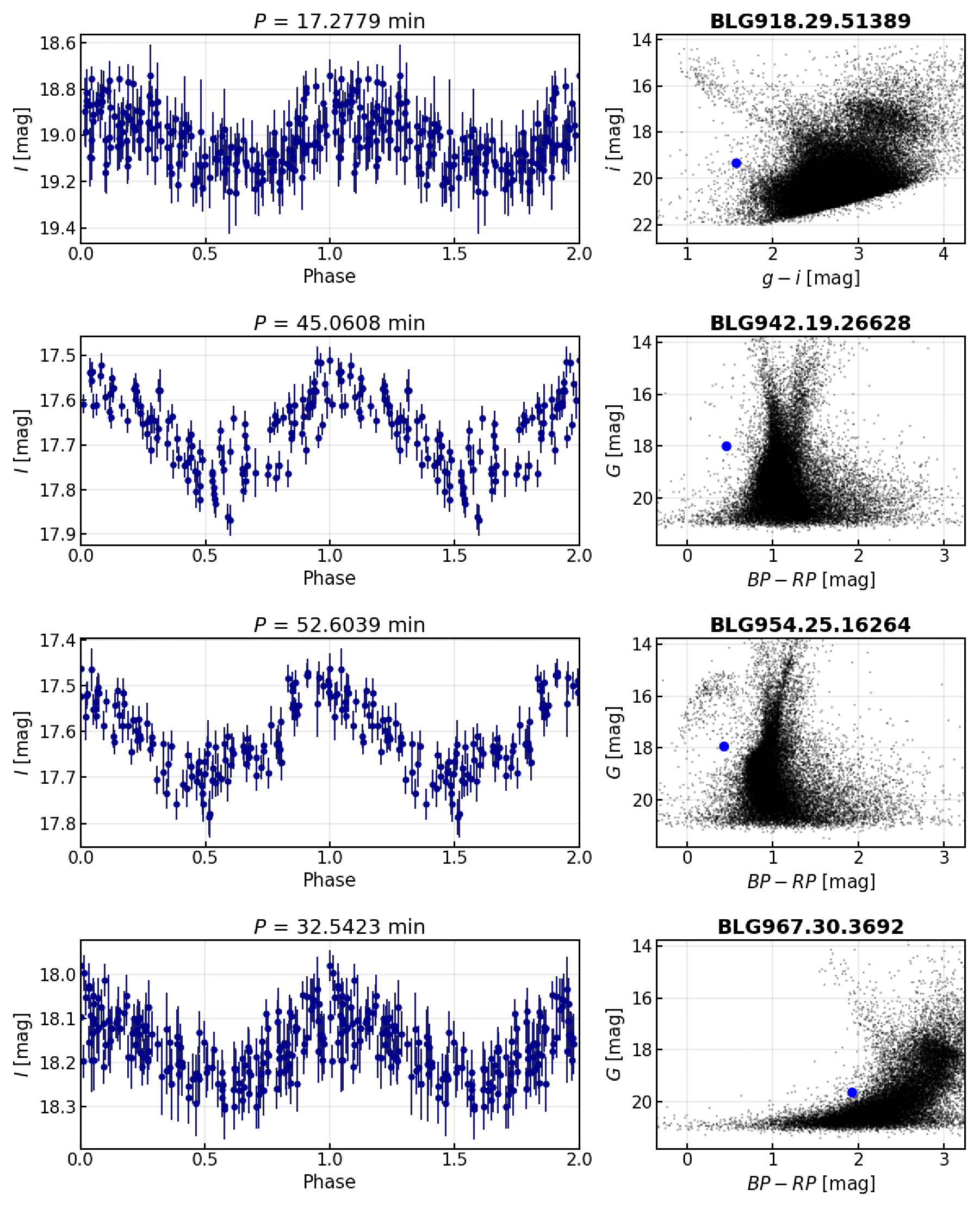}
\FigCap{Same as Fig.~11 for another four objects. The CMDs for BLG918.29.51389 is based on {\it g} and {\it i}-band brightness measurements obtained from the DECam Plane Survey.}
\end{figure}

\begin{figure}
\includegraphics[width=12.71cm]{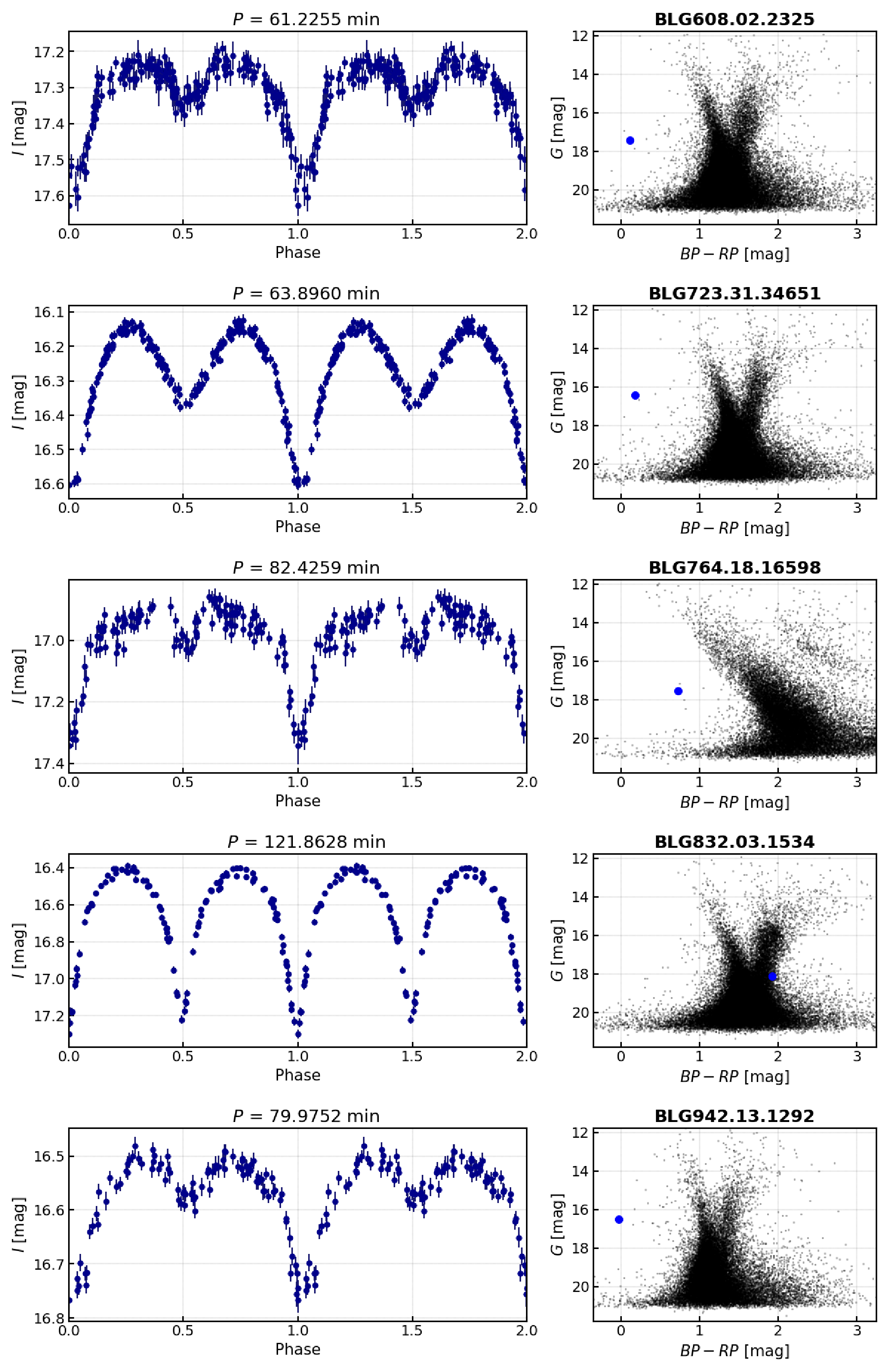}
\FigCap{Phase-folded {\it I}-band light curves (left panels) and CMDs (right panels) for five eclipsing binary systems found in the OGLE-IV outer Galactic bulge fields.}
\end{figure}

\begin{figure}
\includegraphics[width=12.71cm]{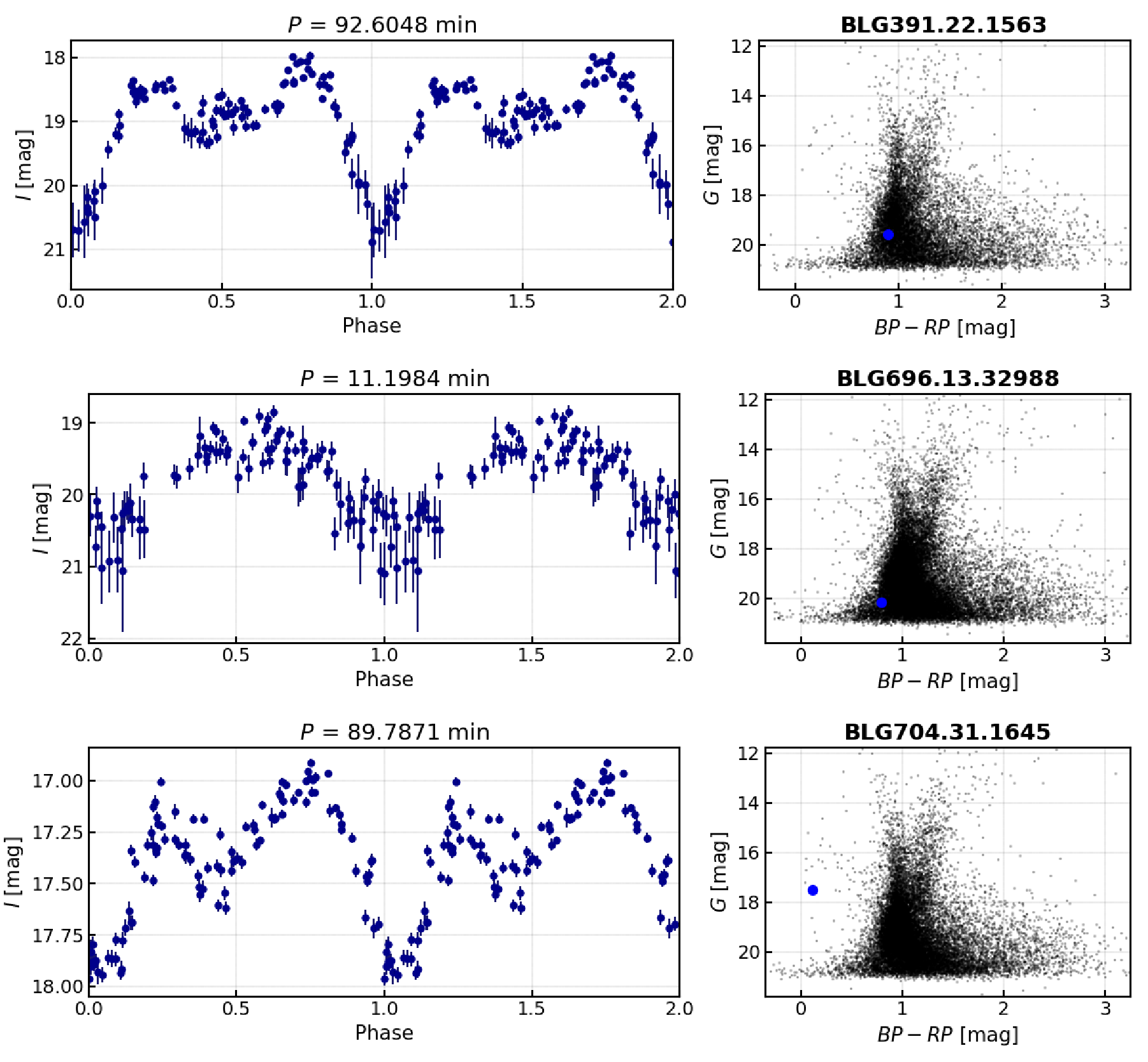}
\FigCap{Phase-folded {\it I}-band light curves and CMDs for three large-amplitude variables of unknown type. Objects BLG391.22.1563 and BLG696.13.32988 have minimum brightness near the detection limit in the shallow outer bulge survey.}
\end{figure}

\begin{sidewaystable}
\caption{Photometric properties of detected BLAPs}
\label{}
{\scriptsize
\begin{tabular}{@{}lllrrccrcllcrc@{}}
\toprule
Name & RA (J2000) & \multicolumn{1}{c}{Dec (J2000)} & \multicolumn{1}{c}{$l$} & \multicolumn{1}{c}{$b$} & $\varpi$ & $\sigma_\varpi$ & \multicolumn{1}{c}{$P_{\rm puls}$} & $\sigma_P$ &  \multicolumn{1}{c}{$< I >$} & \multicolumn{1}{c}{$A_{I}$} & $< G >$ & \multicolumn{1}{r}{$BP-RP$} & Other Name \\ 
& \multicolumn{1}{c}{[${}^\circ$]} & \multicolumn{1}{c}{[${}^\circ$]} & \multicolumn{1}{c}{[${}^\circ$]} & \multicolumn{1}{c}{[${}^\circ$]} & [mas] & [mas] & \multicolumn{1}{c}{[min]} & {\tiny [$ 10^{-7}$ min]} & [mag] & [mag] & [mag] & [mag] &[Ref.] \\ \midrule
OGLE-BLAP-062 & 254.83294 & $-37.11583$ & 347.76988 & 3.34321 & --- & --- & 32.9956511 & 106 & 17.986 & 0.241 & 18.835 & 1.365 & -- \\
OGLE-BLAP-063 & 256.51623 & $-31.67478$ & 352.95038 & 5.54725 & 0.30 & 0.10 & 58.6452979 & 33 & 16.744 & 0.155 & 17.275 & 0.723 & -- \\
OGLE-BLAP-064 & 258.32242 & $-33.58885$ & 352.30587 & 3.18846 & --- & --- & 40.4852661 & 39 & 18.124 & 0.258 & 18.745 & 1.068 & -- \\
OGLE-BLAP-065 & 260.88549 & $-43.96965$ & 344.95315 & $-4.41049$ & 0.07 & 0.18 & 15.1879302 & 7 & 17.817 & 0.292 & 18.365 & 0.771 & -- \\
OGLE-BLAP-066 & 261.29448 & $-39.93427$ & 348.47142 & $-2.39965$ & 0.28 & 0.56 & 27.9819187 & 58 & 18.509 & 0.233 & 19.461 & 1.609 & -- \\
OGLE-BLAP-067 & 261.91435 & $-40.58023$ & 348.19822 & $-3.15237$ & 0.02 & 0.25 & 20.3038847 & 103 & 18.282 & 0.376 & 18.704 & 0.577 & -- \\
OGLE-BLAP-068 & 262.38303 & $-24.60972$ & 1.74050 & 5.34600 & 1.18 & 0.30 & 38.3321045 & 72 & 17.956 & 0.192 & 18.943 & 1.555 & -- \\
OGLE-BLAP-069 & 262.57431 & $-20.65069$ & 5.18119 & 7.34917 & 0.26 & 0.09 & 66.4939433 & 646 & 16.825 & 0.123 & 17.461 & 0.912 & -- \\
OGLE-BLAP-070 & 262.96609 & $-23.06989$ & 3.33021 & 5.73570 & 0.49 & 0.24 & 34.8664728 & 33 & 17.413 & 0.213 & 18.170 & 1.544 & -- \\
OGLE-BLAP-071 & 263.23020 & $-40.16291$ & 349.09850 & $-3.76193$ & 0.29 & 0.05 & 58.4985243 & 30 & 15.113 & 0.154 & 15.577 & 0.648 & -- \\
OGLE-BLAP-072 & 264.47945 & $-18.04700$ & 8.36775 & 7.21315 & 0.10 & 0.09 & 31.8847179 & 36 & 16.976 & 0.326 & 17.169 & 0.391 & -- \\
OGLE-BLAP-073 & 265.45330 & $-20.85852$ & 6.44301 & 4.95813 & --- & --- & 36.0712030 & 39 & 17.397 & 0.264 & 18.126 & 1.073 & -- \\
OGLE-BLAP-074 & 265.56943 & $-16.17125$ & 10.52892 & 7.29922 & 0.14 & 0.15 & 15.3685860 & 11 & 18.057 & 0.286 & 18.177 & 0.184 & -- \\
OGLE-BLAP-075 & 265.65443 & $-38.87139$ & 351.19269 & $-4.65797$ & 0.28 & 0.18 & 22.0857709 & 15 & 18.133 & 0.317 & 18.582 & 0.697 & -- \\
OGLE-BLAP-076 & 267.18553 & $-19.57779$ & 8.38603 & 4.22924 & 0.16 & 0.19 & 34.3470343 & 71 & 17.902 & 0.268 & 18.506 & 1.044 & -- \\
OGLE-BLAP-077 & 268.08081 & $-16.78669$ & 11.23146 & 4.91379 & 0.20 & 0.17 & 10.1121423 & 4 & 17.728 & 0.236 & 18.379 & 0.961 & -- \\
OGLE-BLAP-078 & 268.13777 & $-16.73893$ & 11.30044 & 4.89074 & 0.48 & 0.32 & 22.1350836 & 21 & 18.585 & 0.428 & 19.166 & 1.034 & -- \\
OGLE-BLAP-079 & 269.19956 & $-19.62019$ & 9.31129 & 2.57134 & 0.55 & 0.20 & 41.3926356 & 33 & 17.526 & 0.219 & 18.574 & 1.642 & -- \\
OGLE-BLAP-080 & 269.26293 & $-16.26118$ & 12.25711 & 4.19601 & 0.90 & 0.35 & 21.6268702 & 26 & 18.530 & 0.295 & 19.278 & 1.112 & -- \\
OGLE-BLAP-081 & 269.62850 & $-14.52371$ & 13.94580 & 4.75164 & 0.47 & 0.18 & 38.5197428 & 345 & 17.239 & 0.243 & 18.413 & 1.983 & -- \\
OGLE-BLAP-082 & 270.51874 & $-17.78211$ & 11.52927 & 2.39961 & --- & --- & 53.5543446 & 124 & 17.249 & 0.201 & 17.930 & --- & -- \\
OGLE-BLAP-083 & 275.79400 & $-24.11316$ & 8.32787 & $-4.95472$ & 0.20 & 0.10 & 47.7527083 & 52 & 16.942 & 0.180 & 17.543 & 0.917 & -- \\
OGLE-BLAP-084 & 276.07003 & $-20.11729$ & 11.99588 & $-3.32710$ & --- & --- & 34.2030899 & 32 & 17.654 & 0.264 & 18.393 & 1.099 & -- \\
OGLE-BLAP-085 & 276.25389 & $-24.66896$ & 8.02645 & $-5.58261$ & 0.07 & 0.08 & 46.0804344 & 87 & 16.780 & 0.182 & 17.104 & 0.594 & -- \\
OGLE-BLAP-086 & 276.61851 & $-28.20053$ & 5.01152 & $-7.48189$ & 0.01 & 0.08 & 32.9633868 & 47 & 16.735 & 0.226 & 16.813 & 0.625 & [1] \\
OGLE-BLAP-087 & 277.97158 & $-18.85708$ & 13.94509 & $-4.33657$ & 0.04 & 0.15 & 20.4200231 & 13 & 17.370 & 0.330 & 17.899 & 0.721 & -- \\
OGLE-BLAP-088 & 278.01615 & $-21.00101$ & 12.04936 & $-5.35152$ & 0.33 & 0.21 & 20.5911629 & 8 & 17.873 & 0.361 & 18.208 & 0.423 & -- \\
OGLE-BLAP-089 & 278.01967 & $-19.70939$ & 13.20528 & $-4.76625$ & 0.16 & 0.15 & 25.1626966 & 19 & 17.608 & 0.290 & 18.013 & 0.459 & -- \\
OGLE-BLAP-090 & 278.42267 & $-13.30536$ & 19.08535 & $-2.17429$ & --- & --- & 55.4221598 & 58 & 16.226 & 0.114 & 16.901 & 1.084 & -- \\
OGLE-BLAP-091 & 279.09657 & $-16.54557$ & 16.49809 & $-4.23598$ & 0.54 & 0.03 & 22.5836372 & 52 & 14.767 & 0.124 & 15.105 & 0.400 & -- \\
OGLE-BLAP-092 & 279.74064 & $-22.74435$ & 11.20470 & $-7.56414$ & --- & --- & 7.5140274 & 3 & 17.988 & 0.158 & 17.958 & 0.238 & -- \\
OGLE-BLAP-093 & 281.27844 & $-30.13465$ & 5.03473 & $-11.98657$ & 0.12 & 0.06 & 20.1679637 & 197 & 16.601 & 0.125 & 16.476 & $-0.248$ & [2] \\
OGLE-BLAP-094 & 281.80659 & $-17.13507$ & 17.14000 & $-6.81803$ & 0.21 & 0.06 & 25.3925666 & 26 & 16.496 & 0.357 & 16.777 & 0.339 & -- \\ \bottomrule
\end{tabular}
}
{\scriptsize \\ References: [1] OW J1826-2812 (Ramsay \etal 2022), [2] SMSS J184506.82-300804.7 (Chang \etal 2024)}
\end{sidewaystable}

\begin{sidewaystable}
\caption{Photometric properties of eighteen pulsating-like stars}
\label{}
{\scriptsize
\begin{tabular}{@{}lllrrcclcllcc@{}}
\toprule
Name & RA (J2000) & \multicolumn{1}{c}{Dec (J2000)} & \multicolumn{1}{c}{$l$} & \multicolumn{1}{c}{$b$} & $\varpi$ & $\sigma_\varpi$ & \multicolumn{1}{c}{$P_{\rm puls}$} & $\sigma_P$ &  \multicolumn{1}{c}{$< I >$} & \multicolumn{1}{c}{$A_{I}$} & $< G >$ & \multicolumn{1}{r}{$BP-RP$} \\ 
& \multicolumn{1}{c}{[${}^\circ$]} & \multicolumn{1}{c}{[${}^\circ$]} & \multicolumn{1}{c}{[${}^\circ$]} & \multicolumn{1}{c}{[${}^\circ$]} & [mas] & [mas] & \multicolumn{1}{c}{[min]} & {\tiny [$ 10^{-7}$ min]} & [mag] & [mag] & [mag] & [mag] \\ \midrule
BLG852.09.20745 & 251.16834 & $-29.75587$ & 351.60113 & $10.34601$  & --- & --- & 53.0253606 & 444  & 18.323 & 0.258 & 18.774 & 0.668  \\
BLG481.23.763   & 254.35439 & $-21.53726$ & 0.01575   & $13.16083$  & 0.01  & 0.16 & 47.9596203 & 437  & 17.526 & 0.173 & 18.069 & 0.775  \\
BLG886.28.45247 & 254.79290 & $-32.31533$ & 351.54807 & $6.32243$   & --- & --- & 51.2152787 & 66   & 17.671 & 0.214 & 18.227 & 0.829  \\
BLG882.04.34201 & 254.82191 & $-28.35160$ & 354.73151 & $8.72091$   & 0.15  & 0.17 & 52.3152169 & 670  & 17.582 & 0.173 & 18.016 & ---    \\
BLG887.11.1867  & 254.84920 & $-34.23643$ & 350.05285 & $5.10612$   & --- & --- & 42.7895905 & 57   & 17.218 & 0.104 & 17.971 & 1.080  \\
BLG913.15.4196  & 258.42735 & $-25.00119$ & 359.38741 & $8.10018$   & 0.18  & 0.27 & 53.7033771 & 535  & 18.189 & 0.298 & 19.049 & 1.221  \\
BLG918.29.51389 & 259.04021 & $-32.86844$ & 353.24256 & $3.11955$   & 0.43  & 0.66 & 17.2779357 & 34   & 19.022 & 0.258 & 20.066 & --- \\
BLG807.31.13428 & 262.17954 & $-14.87571$ & 9.91594   & $10.76036$  & 0.23  & 0.13 & 52.4995780 & 493  & 17.257 & 0.157 & 17.882 & 0.870  \\
BLG967.30.3692  & 262.31942 & $-23.61095$ & 2.54932   & $5.94017$   & --- & --- & 32.5423393 & 78   & 18.165 & 0.174 & 19.631 & 1.928  \\
BLG537.24.71551 & 268.79063 & $-37.55382$ & 353.59142 & $-6.10845$  & --- & --- & 51.5481337 & 107  & 18.122 & 0.206 & 18.656 & 0.660  \\
BLG942.19.26628 & 269.79992 & $-41.34727$ & 350.63784 & $-8.64194$ & 1.09 & 0.14 & 45.0608030 & 902  &  17.665 & 0.212 & 18.004 & 0.460  \\
BLG541.09.33260 & 271.46953 & $-38.60803$ & 353.68966 & $-8.46380$  & 0.32  & 0.05 & 47.5686369 & 80   & 15.487 & 0.130 & 15.820 & 0.479  \\
BLG954.25.16264 & 271.92296 & $-43.76137$ & 349.21479 & $-11.15339$ & 0.28  & 0.13 & 52.6039070 & 680  & 17.607 & 0.214 & 17.936 & 0.427  \\
BLG658.19.4003  & 275.32101 & $-31.62993$ & 1.41166   & $-8.03147$  & 0.22  & 0.17 & 49.8625157 & 259  & 17.793 & 0.138 & 18.172 & 0.596  \\
BLG688.16.858   & 275.72691 & $-32.46720$ & 0.81361   & $-8.71653$  & 0.01  & 0.05 & 43.6297893 & 238  & 16.267 & 0.154 & 16.234 & 0.022  \\
BLG382.24.180   & 276.30373 & $-35.99486$ & 357.81897 & $-10.70649$ & 0.29  & 0.05 & 52.7114147 & 3579 & 15.626 & 0.137 & 15.867 & 0.377  \\
BLG555.08.43224 & 276.94071 & $-28.75299$ & 4.64216   & $-7.98335$  & 0.22  & 0.20 & 47.3230106 & 229  & 18.126 & 0.182 & 18.665 & 0.689  \\
BLG399.06.732   & 281.14731 & $-26.74430$ & 8.12033   & $-10.45424$ & 0.16  & 0.06 & 40.1926896 & 288  & 16.519 & 0.103 & 16.494 & $-0.025$ \\ \bottomrule
\end{tabular}
}
\caption{Photometric properties of binary systems and other high-amplitude variables}
\label{}
{\scriptsize
\begin{tabular}{@{}lllrrcclcllcrc@{}}
\toprule
Name & RA (J2000) & \multicolumn{1}{c}{Dec (J2000)} & \multicolumn{1}{c}{$l$} & \multicolumn{1}{c}{$b$} & $\varpi$ & $\sigma_\varpi$ & \multicolumn{1}{c}{$\textit{P}$} & $\sigma_P$ &  \multicolumn{1}{c}{$< I >$} & \multicolumn{1}{c}{$A_{I}$} & $< G >$ & \multicolumn{1}{r}{$BP-RP$} \\ 
& \multicolumn{1}{c}{[${}^\circ$]} & \multicolumn{1}{c}{[${}^\circ$]} & \multicolumn{1}{c}{[${}^\circ$]} & \multicolumn{1}{c}{[${}^\circ$]} & [mas] & [mas] & \multicolumn{1}{c}{[min]} & {\tiny [$ 10^{-7}$ min]} & [mag] & [mag] & [mag] & [mag] \\ \midrule
BLG608.02.2325  & 266.58028 & $-40.91053$ & 349.80731 & $-6.31960$  & 0.48  & 0.09 & 61.2255171  & 38   & 17.309 & 0.354 & 17.425 & 0.117  \\
BLG832.03.1534  & 267.78860 & $-15.99452$ & 11.77647  & $5.55531$  & 5.61  & 0.55 & 121.862758 & 423  & 16.414 & 0.847 & 18.157 & 2.242  \\
BLG942.13.1292  & 269.33622 & $-41.70733$ & 350.14952 & $-8.51156$  & 0.30  & 0.06 & 79.9752372  & 522  & 16.564 & 0.243 & 16.497 & $-0.023$ \\
BLG764.18.16598 & 272.55705 & $-21.64039$ & 9.09754   & $-1.15788$  & 0.62  & 0.11 & 82.4258726  & 107  & 16.979 & 0.457 & 17.543 & 0.723  \\
BLG723.31.34651 & 280.85077 & $-17.02810$ & 16.82668  & $-5.95244$  & 0.44  & 0.06 & 63.8959988  & 21   & 16.269 & 0.467 & 16.440 & 0.184  \\ \midrule
BLG696.13.32988 & 277.79776 & $-34.18815$ & 0.01607   & $-11.02687$ & --- & --- & 11.1984346  & 56   & 19.680 & 1.629 & 20.169 & 0.792  \\
BLG704.31.1645  & 279.19703 & $-34.29750$ & 0.41518   & $-12.12097$ & 0.50  & 0.12 & 89.7870609  & 1014 & 17.337 & 0.908 & 17.516 & 0.118  \\
BLG391.22.1563  & 279.87395 & $-37.19227$ & 357.93805 & $-13.81055$ & 1.04  & 0.43 & 92.6047602  & 3496 & 18.850 & 2.536 & 19.585 & 0.901  \\ \bottomrule
\end{tabular}
}
\end{sidewaystable}

\Acknow{We thank all the OGLE observers for their contribution to the collection of the photometric data over the decades. This work has been funded by the National Science Centre, Poland, grant no.~2022/45/B/ST9/00243 to I.S. For the purpose of Open Access, the author has applied a CC-BY public copyright licence to any Author Accepted Manuscript (AAM) version arising from this submission. We used data from the European Space Agency (ESA) mission Gaia, processed by the Gaia Data Processing and Analysis Consortium (DPAC). Funding for the DPAC has been provided by national institutions, in particular the institutions participating in the Gaia Multilateral Agreement.}

\end{document}